# TRACING THE DYNAMICS OF DISK GALAXIES WITH OPTICAL AND INFRARED SURFACE PHOTOMETRY: COLOR GRADIENTS IN M99

Rosa A. González[1,2,4] and James R. Graham[2,3,4]

Astronomy Department, University of California, Berkeley CA 94720-3411

astro-ph/9511025  6 Nov 1995

---

[4]e-mail: ragl@astron.berkeley.edu,jrg@astron.berkeley.edu

## ABSTRACT

We present optical and IR surface photometry of M99 (NGC 4254) at $g$, $r_S$, $i$, $J$ and $K'$. We also present a $K'$ image of M51 (NGC 5194) for comparison. We use these data to investigate if the presence and strength of spiral structure are wavelength dependent, and hence if they are features in the old stellar disk or in the gas and dust only or associated with star formation or some combination. The arms of M99 are still very prominent in the infrared, although broader and smoother than at optical wavelengths. Fourier decomposition of the disk light reveals that the radial distribution of power depends on wavelength, which in turn implies that the spiral structure traced in the visual (i.e. young population I and dust) is different from the one detected at 2 $\mu$m (i.e. old stellar disk). While M51 has significantly more power in $m = 2$ than in any other mode at all radii, M99 shows noticeable power at $m = 2$, $m = 1$ and $m = 3$. We observe radial modulation of the power and a dependency of power with wavelength that are consistent with modal theory of spiral structure.

A central motivation for our research is the fundamental idea of density wave theory that the passage of a spiral density wave triggers star formation as it shocks the molecular gas in the disk. If this is the case color gradients, due to stellar population age gradients, should occur across the spiral arms. We have found a gradient consistent with this scenario in a reddening-free $Q$-like parameter at 6 kpc galactocentric distance across one of the arms of M99; this photometric parameter, $Q_{r_SJgi}$, is very sensitive to red supergiants. We rule out that the change in $Q_{r_SJgi}$ across the arm is mainly due to dust. We also see a difference in $Q_{r_SJgi}$ going from the interarm regions to the arms that indicates that arms cannot be due exclusively to crowding of stellar orbits.

We present the first measurement of $\Omega_p$, the angular speed of the spiral pattern, and of the location of the corotation radius, derived from the drift velocity of the young stars away from their birth site. The measured $Q_{r_S J g i}$ implies a star formation rate for M99 within the range of 10–20 $M_\odot \text{yr}^{-1}$; a disk stellar mass surface density of $\simeq 80$ $M_\odot \text{pc}^{-2}$; and a maximum contribution of $\simeq 20$ percent from red supergiants to the $K'$ light in a small region, and much smaller on average. We measure a $K'$ arm–interarm contrast of 2-3, too high for M99 to be a truly isolated galaxy.

*Subject headings:* galaxies:individual (M51) – galaxies:individual (M99) – galaxies:infrared – galaxies:kinematics and dynamics – galaxies:photometry – galaxies:spirals

## 1. Introduction.

The morphology of a spiral galaxy is of great importance because it is often the only available clue to the galaxy's underlying structure and dynamics. However, the morphology of spirals is highly dependent on the wavelength at which they are observed; the IR morphology is known for only a few spirals, and in several of these cases the IR and optical morphologies differ significantly – i.e. in the presence or absence of bars, the degree of flocculence of the arms, even the apparent Hubble type (Hackwell & Schweizer 1983; Scoville et al. 1988; Thronson et al. 1989; Block & Wainscoat 1991; Block et al. 1994). Past studies of spirals have used optical surface photometry to probe the dynamics of stellar disks (e.g. Schweizer 1976; Talbot, Jensen, & Dufour 1979; Elmegreen 1981; Iye et al. 1982; Elmegreen & Elmegreen 1984, 1985, 1990; Elmegreen, Elmegreen, & Seiden 1989; Elmegreen, Elmegreen, & Montenegro 1992). However, even at $I$ the observed morphology of disk galaxies is strongly influenced by extinction. Combined optical and near infrared data extending to a wavelength of 2 $\mu$m offer a better opportunity for measuring extinction and reddening. Correcting for extinction is crucial when using morphology as a probe of disk dynamics, since reddening and extinction can lead to incorrect estimates of the contribution of young stars and the true arm–interarm density contrast. Also, given that the effect of dust extinction is much less severe in the IR relative to the optical ($A_K = 0.08 A_B$), the IR surface brightness distribution is a more direct measure of the stellar disk mass surface density distribution (Rix & Rieke 1993).

We present the first results of an ongoing optical–IR survey of $\sim 25$ spiral galaxies of different types and a range of inclinations. The primary goal of our study is to discover the nature and triggering mechanisms of spiral structure.

Modal theory (Bertin et al. 1989a,b), for example, predicts that only lower order modes ($m = 1$ and $m = 2$) will survive in an old stellar disk. Conversely, swing amplification (Toomre 1990; Toomre & Kalnajs 1991) predicts that gas should form a stronger and more tightly wound spiral than the old stellar population.

The Sc I galaxy M99 occupies a special place in our survey. Its spiral pattern appears to be exceptionally strong (Schweizer 1976); it has an unusual three–arm spiral with a very prominent $m = 1$ mode, i.e. one arm is much stronger than the other two. In the optical, its arms show a hint of the azimuthally asymmetric profiles that would be produced if star formation is induced by the passage of a spiral density wave (Schweizer 1976). Assuming a distance of 16 Mpc (Jacoby et al. 1992), M99 offers good spatial resolution ($1'' = 78$ pc); the galaxy has $R_{25} = 2'.69$ (Elmegreen et al. 1992).

M51 (NGC 5194) is one of the most regular two–armed "grand design" spirals in the sky. (Type SbcI, distance = 9.6 Mpc, for 46.5 pc/$''$ (Schweizer 1976); $R_{25} = 5'.48$ (Elmegreen et al. 1992).) The $K'$ data of M51 has allowed interesting comparisons with M99 regarding their azimuthal distribution of light at that wavelength.

## 2. Observations.

Figure 1 shows M99 at $g$ ($\lambda = 5000$Å). The optical images were obtained in 1994 April 15 and 16 UT with the 1-m Nickel telescope at Lick Observatory. The pixel scale is $0''.37$/pixel, for a field of view of the Ford CCD $2048^2$ (binned $2 \times 2$) of $6'.3$. Given the large size of the galaxy, usually we did not dither more than a fraction of an arcminute between each (typically 300 to 900 seconds of time) individual exposure. Frames were bias- and dark current-subtracted prior to flat-fielding with twilight flats. Bad columns and pixels were masked out, and a constant was fit to the object-free areas for sky subtraction. Individual frames were registered to the nearest pixel (i.e. without any interpolation). Cosmic rays were identified by subtracting a median image from each shifted individual frame and then masked out. The frames were then coadded and the resulting mosaic was divided by an exposure map, thus getting units of Analog-to-Digital Units (ADU) s$^{-1}$ at every pixel. A final measure of sky was performed at the edges of the mosaic. Since the galaxy extends almost all the way to the edges of the chip, we expect that the biggest systematic error in the photometry will come from overestimating the brightness of the sky: at 0.9 $R_{25}$, the galaxy surface brightness is 3 percent of the sky brightness in the $g$-band and 1.5 percent of sky in the $i$-band.

The photometric calibration in the optical bands was done in the Thuan-Gunn system (Thuan & Gunn 1976; Wade, Hoessel, & Elias 1979; Kent 1985). This system is calibrated with the primary standard star BD + 17°4708 (sd F6). Since we did not observe BD + 17°4708 in April, we calculated synthetic Thuan-Gunn magnitudes for the spectrophotometric standard star Feige 34 in our three bandpasses. Each bandpass was defined as the product of the respective filter transmission curve, the Q.E. curve of the Ford CCD we used and the atmospheric transmission curve for Lick from Hayes (1970); the fluxes of BD + 17°4708 and Feige 34 were taken from Oke & Gunn (1983), Massey et al. (1988), and Massey & Gronwall (1990). These synthetic magnitudes were then used to assign empirical magnitudes to the other standard stars observed in April 15 and to find the zero-point of that night. The night of April 16 was not photometric, but we calibrated the final mosaics by measuring stars in the frames taken on the 15$^{th}$.

Figure 2 shows M99 at $K'$ ($\lambda = 2.16 \mu m$); the arms are marked following the nomenclature of Iye et al. (1982). The infrared images were obtained in 1994 March 14 UT with the 1.3-m telescope at Kitt Peak; the Infrared Imager (IRIM) uses a 256x256 NICMOS3 detector and has a field of view of $8'.5$ (Probst 1993; Heim, Buchholz, & Luce 1994). We obtained 2 min frames (coadditions of 30 s exposures at $J$ and 20 s exposures at $K'$), sequentially positioning the galaxy in the four quadrants of the array and nodding to sky only every fifth frame. The first step in the reduction was a pixel by pixel linearization. The images were then flatfielded, $J$ with skyflats and $K'$ with domeflats. Sky frames were multiplicatively scaled to the value at the edges of object frames and subtracted from them (sky subtraction takes care also of the dark current); the sky frames were made by medianing (i.e. averaging the two central values for each pixel) the four frames adjacent to each galaxy frame, after masking out bright stars and extended objects. Individual frames were then sub-pixelated $\times$ 4 in both dimensions and registered to the nearest $0''.5$, once again without any additional interpolation. The rest of the reduction was analogous to that of the optical images. Despite the large size of the final IR mosaic, sky subtraction will still be the biggest contribution to the systematic error in the photometry (estimated to be 1–2 percent).

The photometric calibration of the IR images was done with faint UKIRT standard stars (Casali & Hawarden 1992). For the calculation of the zero point, we assumed $K' = K + 0.2(H - K)$ (Wainscoat & Cowie 1992).

Figure 3 shows the $K'$-band image of M51. It was obtained in 1994 March 15 at Kitt Peak, under non-photometric conditions. Given the angular size of M51 we offset frequently to sky. A field of view smaller than the object and the changing sky in cloudy conditions add up to a bigger uncertainty in the sky subtraction; however, we used rms-minimization to improve the scaling of sky to object (i.e. in each case we used the sky frame that minimized the noise in the sky-subtracted image; the sky-subtracted frame with the lowest rms was deemed the reference image; and finally we added a constant to each of the remaining images to shift the level at the edges to the value at the same location in the reference image). Our final mosaic is bigger and therefore includes regions farther away from the center of the galaxy than recently published data (Rix & Rieke 1993); we should then have a better estimate of the sky level.

Table 1 gives a summary of the observations.

## 3. Analysis.

All the images were co-registered, using the $r_S$ image as a reference, and deprojected to a face-on view (Fig. 4). For deprojection, we first rotated the galaxy to make the major axis parallel to the columns (i.e. rotated by an angle equal to the negative of the PA); then, we stretched the pixels along rows (i.e. the minor axis) by a factor equal to $1/\cos(\omega)$, where $\omega$ is the inclination angle. For M99, we used PA = 68°, $\omega$ = 42° (Phookun, Vogel, & Mundy 1993). We decided to use these numbers for consistency, since we use the rotation curve of M99 published in this reference. However, it is worth mentioning that although Phookun et al. used the procedure ROTCUR (Begeman 1989) to simultaneously determine the rotation curve, PA and $\omega$, they forced the two latter parameters to remain close to the values that maximize the power in the axisymmetric component (PA = 70°, $\omega$ = 42°, Iye et al. 1982). It is not possible to determine kinematical inclinations for galaxies with $\omega < 40°$ (Begeman 1989) and M99 must be very close to that limit, as judged from the numbers used by Schweizer (PA = 58°, $\omega$ = 29°). With the exception of mode analysis, our conclusions should not be significantly affected by the choice of PA and $\omega$. For M51 we took both PA and $\omega$ from Schweizer (PA = $-10°$, $\omega = 20°$).

Next, we unwrapped the spiral arms by plotting the images in a $\theta$ vs. $logR$ map (Fig. 5) (cf. Iye et al. 1982; Elmegreen et al. 1992). Under this geometric transformation, logarithmic spirals will appear as straight lines with slope = $\ln(10) \times \cot(-i)$, where $i$ is the pitch angle if the azimuthal angle increases in the direction of rotation and the spiral arms trail. These unwound or unwrapped images are useful to measure the pitch angle of the arms, to measure the arm–interarm contrast by taking cuts along columns, and to calculate the Fourier power spectrum at each radius. Our program allows us to choose arbitrary smoothing lengths in both $\theta$ and $R$, and the output images are in units of ADU s$^{-1}$ $\square''^{-1}$. For example, it is easy to produce a radial light profile simply by smoothing in $\theta$ with an angle equal to $2\pi$.

To reduce the contamination by point sources (i.e. foreground stars and HII regions) in the unwrapped images, we first masked pixels that were $10\sigma$ or more above the average noise of smooth looking arm regions in the unsharp-masked $g$ image, and then used the median surface brightness. We smoothed the data in bins with a radial extent of $r_{min}/r_{max} = 1.049$ and an azimuthal extent of $4°.875$; the separation between bins is $r_2/r_1 = 1.008$ in radius and $1°.2$ in azimuth. These smoothing lengths are similar to those used in previous studies (Schweizer 1976; Rix & Rieke 1993) and allow direct comparisons of results. At the same time, the use of bin spacings smaller than the smoothing lengths – equivalent to boxcar smoothing – reveals structure at finer scales than the aforementioned works. For the error analysis, we smoothed the images in disjoint bins (which are statistically independent) and used smoothing lengths of the order of the spatial resolution of the data.

## 4. Results.

### 4.1. Mode Analysis.

Using the unwound images, we calculated row by row power spectra for M51 and M99. We define the power in each mode $m > 0$, at each radius, as

$$P(m) = 2\frac{|H(m)|^2}{|H(0)|^2}. \tag{1}$$

Here,

$$H(m) = \int_0^{2\pi} I(\theta)e^{2\pi i m\theta} d\theta, \tag{2}$$

with $I(\theta)$ being the surface brightness; we are implicitly integrating over all pitch angles.

The interpretation of Fourier transforms as a probe of spiral structure can be complicated by contamination by foreground stars, sky gradients and internal extinction (Elmegreen, Elmegreen, & Montenegro 1993); also, the distribution of power between different modes is highly sensitive to the deprojection parameters. Moreover, it is hard to estimate errors in the power spectra because pixels are no longer statistically independent after taking the Fourier transform. However, both the fact that at all wavelengths – and in particular at $g$ and at $K'$ (Fig. 7a and 7b) – we seem to recover similar information about M99 and the small "false alarm probabilities" of peaks in periodograms of the same data (cf. Press et al. 1992) give us confidence in the following general conclusions.

We confirm that, at all radii (Figure 6), M51 has significantly more power in $m = 2$ than in any other mode (Rix & Rieke 1993). Although M99 (Fig. 7) displays a dominant $m = 2$ mode between 0.3 and 0.5 $R_{25}$, it shows significant power at $m = 1$ and $m = 3$ too, with $m = 1$ stronger inside $\sim 0.45 R_{25}$ and $m = 3$ peaking outside this radius. The significant strength of $m = 1$ is consistent with a non-linear interaction between it and $m = 2$ exciting $m = 3$ (Bertin 1995); a driving of $m = 3$ by $m = 2$ has also been suggested by Elmegreen et al. (1992). The ripples of $m = 2$ are likely real, since we observe them both at $g$ and at $K'$; it can be explained as the result of interference between the waves that support the mode (Binney & Tremaine 1987; Bertin 1993). Finally, while $m = 2$ shows comparable strength at $g$ and at $K'$ (Fig. 7), there is a trend for higher order modes, i.e. $m = 4$ and $m = 5$, to get fainter at longer wavelengths relative to $m = 0$ (Fig. 8a and 8b); this is consistent with modal theory, which predicts that higher order modes will be stronger in the gas and dust than in the stellar disk since, as stars age, their velocity dispersion increases and smooths out small scale features.

### 4.2. Azimuthal Color Gradients.

#### 4.2.1. A Red-Supergiant-Sensitive Q-Parameter.

If the passage of a spiral density wave and the consequent shocking of molecular clouds triggers the formation of stars, then the mean age $t_{age}$ of the young stellar population is related to the distance $d$ from the arm, at a fixed radius, through the difference between the angular rotation speed, $\Omega(R)$, and the pattern speed, $\Omega_p$, i.e. $d = [\Omega(R) - \Omega_p]Rt_{age}$. In this case, stellar population synthesis models (Charlot & Bruzual 1991; Bruzual & Charlot 1993) predict that the color evolution across an arm is dramatic enough that age gradients can be spatially resolved in a nearby spiral even if the ratio of young to old stars is as small as 0.5 percent by mass. Figure 9 shows the model $Q$-like, reddening-free parameter $Q_{RJVI} = (R-J) - \frac{E(R-J)}{E(V-I)}(V-I) = (R-J) - 0.90(V-I)$ (Mihalas & Binney 1981) vs. $t_{age}$, for an instantaneous burst superimposed on a background of 98 percent of old (age $5 \times 10^9$ yr) stars (see §4.2.2); the young population has a Salpeter IMF with $M_{lower} = 0.1 M_\odot$ and $M_{upper} = 100 M_\odot$. $Q$ is reddening insensitive for screen absorption assuming the extinction curve of Rieke & Lebofsky (1985).

$Q$ is an excellent tracer of star formation, since it has almost the same value for all except the reddest stars (Figure 10). Writing $Q$ in a slightly different way helps to appreciate this fact:

$$Q_{RJVI} = log_{10} \frac{I_V^{2.25} I_J^{2.50}}{I_R^{2.50} I_I^{2.25}} \qquad (3)$$

A combination of small numerator and large denominator in the quotient (i.e. low $V$ and $J$ surface brightnesses, and high $R$ and $I$ fluxes) will yield a small and even negative $Q$; a large numerator plus small denominator (i.e. high $V$ and $J$ surface brightnesses, and low $R$ and $I$ fluxes) will result in a high $Q$. It is easy to achieve a low value of $Q$ with a single Planckian peaking around 7000Å (Fig. 11a); however, to obtain a high value of $Q$ it is necessary to have a double-humped spectral energy distribution with a valley around 7000Å, and increasing intensity both redward and blueward of that wavelength (Fig. 11b). Therefore, $Q$ will have a higher value for a mixture of blue and red stars than for just about any single star. Extreme values of $Q$ occur when the most massive stars of the young population become red supergiants. Indeed, the plot in Figure 9 displays a very pronounced peak $\sim 1.5 \times 10^7$ yr after the starburst due to the appearance of luminous, red stars; it also shows that $Q$ is indeed reddening-insensitive for screen absorption and negligibly affected by a mix of dust and stars with plane-parallel geometry and $\tau_V = 1$ or $\tau_V = 2$ in a galaxy with $\cos(\omega) = 0.68$, which is slightly more edge-on than M99 (Bruzual, Magris-C., & Calvet 1988).

In general, extinction and reddening properties are highly dependent on geometry. The widely used Witt, Thronson, & Capuano (1992) models of radiative transfer within galaxies have spherical symmetry, and consist of mixtures of stars and dust with different relative spatial distributions. Figure 12 shows $Q_{RJVI}$ for the Witt et al. "starburst galaxy" model with $\tau_V$=1–10; $Q$ is no longer reddening-insensitive for $\tau_V > 2$. Their "dusty galaxy model" offers a similar picture, with $Q_{RJVI}$ changing more than our error for $\tau_V > 3$. These are the most unfavorable (i.e. optically thick) Witt et al. models that apply to a galactic disk and it is unlikely that $\tau_V$ is $> 2$ for a face-on galaxy (Bruzual et al. 1988; Peletier & Willner 1992; Peletier et al. 1995), so $Q$ should be reddening-insensitive for the disk of M99. Notice that $Q$ is bigger at larger optical depths.

Figures 13 and 14 show $Q$ vs. $t_{age}$ for our set of filters – i.e. $Q_{r_SJgi} = (r_S - J) - \frac{E(r_S-J)}{E(g-i)}(g - i) = (r_S - J) - 0.82(g - i)$, using the extinction curves of Schneider, Gunn, & Hoessel (1983), and Rieke & Lebofsky (1985) –, for different $M_{upper}$, different durations of the starburst, and different percentages of young stars.

Figure 15 displays an unwound $Q_{r_SJgi}$ image of M99. This image compellingly shows $Q$ differences between the arms and the interarm regions and, more importantly, coherent $Q$ changes through the arms. A previous study applying similar techniques to $UBV$ surface photometry of M83 (Talbot et al. 1979) had detected regions either with or without star formation, but no intermediate color. We have focused our attention on a patch in the N1 arm, at $\sim 5.9$ kpc from the center of the galaxy, that from the $J - K'$ image (Figure 16) appears to have little contamination by dust ($J - K'$ is a good dust tracer and suffers less contamination from young stars than, say, $g - K'$). Even column by column (i.e. in annuli having width equal to the spatial resolution of the data), this spot exhibits an azimuthally asymmetric profile in $Q_{r_S J g i}$ that is consistent with coherent star formation induced by the passage of the spiral density wave; to increase the signal-to-noise, we have collapsed the patch from $r \sim 60''$ to $\sim 95''$. Figure 17 shows, superimposed, the $Q_{r_S J g i}$ profile and the $J - K'$ profile of this patch. In this region, $Q_{r_S J g i}$ increases even as the amount of dust, inferred from $J - K'$, starts to decrease away from the dust lane. This is the opposite of what the Bruzual et al. and Witt et al. models of radiative transfer in mixed stars and dust predict for a diminishing optical depth, and therefore we can rule out that the change in $Q_{r_S J g i}$ is due mainly to dust. Figure 18 shows a full azimuthal cut of $Q_{r_S J g i}$ at $R \sim 75''$, as well as the $K'$ azimuthal light profile for the same radii. $Q_{r_S J g i}$ changes at the location of every arm, and the arm $Q$-profiles are more similar to each other than the arm light profiles among themselves; a similar process seems to be at work in the three arms that is compatible with a change of age of the stars across them. If arms were due to crowding of stellar orbits exclusively we should not see any change in the $Q$-parameter going from the interarm regions to the arms.

### 4.2.2. The Stellar Drift and the Pattern Speed.

Figure 19 shows the $Q_{r_S J g i}$ profile vs. distance from the arm for the region marked in Fig. 15 and 16, and two model $Q_{r_S J g i}$ curves with 1 and 2 percent by mass of young stars, respectively. Both models have been calculated with a Salpeter IMF with $M_{lower} = 0.1 M_\odot$ and $M_{upper} = 10 M_\odot$, and a duration of the burst of $2 \times 10^7$ yr – past studies on OB associations, both galactic and extragalactic, suggest that star formation proceeds in them for $1 - 2 \times 10^7$ yr (Elmegreen & Lada 1977; Doom, De Greve, & de Loore 1985; Massey et al. 1989; Regan & Wilson 1993; Gruendl 1995). In both models the young burst is superimposed on a background of stars $5 \times 10^9$ yr old. The choice of 1–2 percent of young population seems reasonable compared both to the number of times that the region under consideration has encountered the spiral pattern (once every $\sim 10^8$ yr) in the lifetime of the disk ($\sim 10^{10}$ yr) and to the amount of light in the arms contributed by old stars in the $B$ band as estimated by Schweizer (1976). It is also not that surprising that the data is well fit by a young population lacking extremely massive stars because we have masked out the brightest stars and the HII regions.

We have chosen the origin of the burst, $d = 0$, at the location of the dust lane in the data, i.e. the location of the shock (Roberts, Huntley, & van Albada 1979). To transform the time in the models into distance from the site of star formation we have adopted the flat rotation curve with $V_{rot} = 140$ km s$^{-1}$ derived by Phookun et al. (1993) from HI data, and then chosen the difference between $\Omega(5.9 \text{ kpc})$ and $\Omega_p$ that allows a good fit of the model $Q_{r_SJgi}$ to the data. Taking $\Omega(5.9 \text{ kpc}) - \Omega_p = 6.5$–$8.0$ km s$^{-1}$ kpc$^{-1}$ yields $\Omega_p = 17.2$–$15.7$ km s$^{-1}$ kpc$^{-1}$, and places $R_{CR}$ at 0.6–0.7 $R_{25}$. This is slightly larger than the 0.54 $R_{25}$ found by Elmegreen et al. (1992), but in better agreement with modal theory of spiral structure (Bertin et al. 1989a), which predicts that the corotation zone is at $\sim 3$ radial scale lengths, i.e. at a location where the gaseous mass is plentiful enough to play a significant dynamical role. (By fitting the radial light profile of M99 between 59″ and 111″, we have measured a scale length that varies between 0.24 $R_{25}$ at $g$ and 0.20 $R_{25}$ at $K'$.)

### 4.2.3. The Star Formation Rate.

By interpreting the observed $K'$ flux in the interarm region downstream of the shock in terms of a population synthesis model with the same parameters used for $Q$ (a mixture of 1–2 percent of stars $\sim 2 \times 10^8$ yr old, estimated from their distance from the dust lane, and 99–98 percent of stars $5 \times 10^9$ yr old), we find a disk stellar mass surface density of 80–75 $M_\odot \text{pc}^{-2}$. With 1 percent of young stars formed in $2 \times 10^7$ yr these numbers imply a star formation rate of 10 $M_\odot \text{yr}^{-1}$, if integrated over the area inside 0.7 $R_{25}$, and 20 $M_\odot \text{yr}^{-1}$, if all the area inside $R_{25}$ is included. This result is in the range of 0–20 $M_\odot \text{yr}^{-1}$ found by Kennicutt (1983) for Sc galaxies. It is important to note, though, that he derived his result from $H\alpha$ flux and $UBV$ colors, i.e. mainly from the output of stars more massive than 10 $M_\odot$, which make up $\sim 6$ percent of the mass of all stars; we are sampling stars less massive that 10 $M_\odot$ and therefore a much bigger fraction of the newly formed stars. Combined with a detected color gradient that is consistent with the evolution in time of a stellar population whose birth was triggered at the location of the dust lane, this general agreement suggests that at least some of the star formation in M99 is associated with the spiral density wave.

### 4.3. Arm–Interarm Contrast.

We measure arm–interarm intensity contrasts by taking cuts along columns in the unwrapped images. Arm–interarm contrasts at different wavelengths have been considered a good discriminant between arm triggering mechanisms. For example, flocculent arms due to self-propagating star formation would show large azimuthal intensity contrasts in the blue – which traces dust and hot, young stars – and very small ones in the infrared – which traces the mass density variations in the disk (Elmegreen & Elmegreen 1984; Rix & Rieke 1993). Following Elmegreen & Elmegreen, we define arm–interarm contrast as the ratio of an intensity maximum in an azimuthal cut to the average of the intensity of the two adjacent minima.

Figures 20a and 20b show azimuthal brightness profiles of M99 in $g$ and $K'$, and in $r_S$ and $K'$, respectively, at $R = 75''$. At optical wavelengths the contrast appears quite altered due to the presence of dust, as suggested by direct images and by dust tracers like $J - K'$; we hence conclude that it is difficult to measure intrinsic arm–interarm contrasts in visible bands. When estimated at $K'$, the arm–interarm contrast of M99 rises from $\simeq 2$ at $30''$ (0.2 $R_{25}$) to $\simeq 3$ at $80''$ (0.5 $R_{25}$), then falls back again to a factor of 2 at 0.6 $R_{25}$; it then shows an abrupt jump to $\simeq$ 4–5 between 0.8 and 0.9 $R_{25}$. For M51, the $K'$ arm–interarm contrast goes from 1.7 at $70''$ (0.2 $R_{25}$) to 2.6 at $125''$ (0.4 $R_{25}$), then declines steadily to a factor of 1.4 at 0.7 $R_{25}$. These numbers are in excellent agreement with Rix and Rieke (1993). Thus, for the range of radii where we can carry out comparisons the $K'$ arm–interarm contrast of the two galaxies is very similar.

At 6400Å, Schweizer (1976) measured an "arm strength" for M99 that was almost twice that for M51; he defined arm strength as the ratio of "arm" light to "disk" light. At each radius, "disk" meant all the light below the mean surface brightness of the two deepest minima separated by more than 90 degrees in azimuth, "arms" were defined as the average amount of light above the "disk". Although Schweizer tried to avoid dust lanes in the choice of minima, the $K'$-band is less affected by dust than optical wavelengths and therefore better for this sort of experiment; also, the arm–interarm contrast is likely a more objective estimator of a galaxy's mass spatial distribution than the "arm strength", since it does not involve interpolating across the arms to determine the old disk surface brightness. As already stated above, from the point of view of their $K'$-band arm–interarm contrast, M99 and M51 seem to be similar objects.

Since we have used the $K'$-band to estimate the arm–interarm contrast (as well as the star formation rate), it is important to rule out a large contribution from red supergiants to the $K'$ light, i.e. that instead of actually tracing mass $K'$ is seeing a few young red stars with a low $M/L$ ratio. In the case of M51, after masking out HII regions and point sources, Rix & Rieke (1993) inferred from the strength of the CO (2.3 $\mu$m) molecular absorption that red supergiants contributed more than 20 percent to the $K$ (2.2 $\mu$m) light only in a small localized region of one of the arms. Using the CO index $(K - CO)$, Rhoads (1995) found for NGC 1309 that, while young cool supergiants do in fact dominate the 2 $\mu$m light of active star forming regions – the bright HII regions we masked out –, when the smooth areas of the spiral arms are included the measured CO index converges to a value consistent with a population $\sim 6 \times 10^8$ yr old (the most massive stars of such a population would be 3–4 $M_\odot$: no supergiants!); unfortunately, Rhoads does not show the index for the smooth arm regions alone. Our results are consistent with the ones just quoted. The fit to the observed $Q_{r_S J g i}$ with 1–2 percent young stars by mass and an upper mass limit of the burst of 10 $M_\odot$ implies that less than 0.1 percent by mass of red supergiants should be present, and their peak contribution to the $K'$ light is 15–25 percent between the ages of $4.5 \times 10^7$ and $10^8$ yr. This is a short interval, and on average the red supergiant contribution will be much less.

Discarding a bigger contribution from red supergiants to $K'$ has significant consequences for disk dynamics. An arm–interarm mass contrast $\geq 2$ is too big to be the result of a long-lived spiral density wave in an isolated disk, i.e. in the absence of driving by a bar or companion, or of a disk cooling mechanism like gas infall (Sellwood & Carlberg 1984; Sellwood 1995). In the case of M99, therefore, the size of the $K'$ arm–interarm contrast is an indication that the galaxy is only apparently isolated. Indeed, HI data shows signatures of gas infall, namely clouds superposed on and beyond the HI disk, at velocities up to 150 km s$^{-1}$ from those found for the disk gas (Phookun et al. 1993), and excess emission in one of the high velocity wings of the double-horned HI profile observed with the Arecibo 305-m telescope (Shulman, Bregman, & Roberts 1994).

## 5. Summary and Conclusions.

As big format infrared detectors have become increasingly available, observations of disk galaxies have revealed spiral arms that, although broader and indeed smoother than in the optical, have turned out to be much stronger and better defined than foreseen, even at $K'$ (Block & Wainscoat 1991; Block et al. 1994; Rix & Zaritsky 1995). It does appear that with the much reduced obscuration by dust the intrinsic structure of galactic disks is finally being unveiled and, with it, the signatures of the underlying dynamics. Resulting from the combination of arm strength in the infrared and better detectors, the signal-to-noise of our data is significantly superior to previous observations.

Fourier decomposition of the disk light of M99 reveals that the radial distribution of power depends on wavelength, which in turn implies that the spiral structure traced in the visual (i.e. young population I and dust) is different from the one detected at 2 $\mu$m (i.e. old stellar disk). The radial modulation of the power and the dependency of power with wavelength that we observe are consistent with modal theory of spiral structure.

We have identified a $Q$-like photometric parameter that is very sensitive to red supergiants. Thanks to very deep optical and near infrared images, we have a positive detection of an azimuthal gradient in $Q_{r_S J g i}$ due to a stellar population *age* gradient across a spiral arm of M99. From the drift velocity of the young stars away from their birth site, we have measured the angular speed $\Omega_p$ of the spiral pattern and the location of the corotation radius. The measured $Q_{r_SJgi}$ implies a star formation rate for M99 within the range of 10–20 $M_\odot\mathrm{yr}^{-1}$; a disk stellar mass surface density of $\simeq 80\ M_\odot\mathrm{pc}^{-2}$; and a maximum contribution of $\simeq 20$ percent from red supergiants to the $K'$ light in a small region and much less on average. According to the models that better resemble our data, spiral structure is both the product of a density wave that involves the whole stellar disk (hence the close to 100 percent by mass of old stars in the arm) and the cause of organized star formation as the wave encounters and shocks molecular gas.

Without very useful discussions with Joan Najita and Arjun Dey, this project might have never been started. We are grateful to Giuseppe Bertin for his comments. Gustavo Bruzual and Stéphane Charlot modified their programs to better suit our needs. Ron Probst, Nick Buchholz and Mike Merrill provided invaluable assistance for observing with IRIM. The Lick Observatory staff was always there when we needed them. We acknowledge the financial support received from the Sloan Foundation and from the Packard Foundation. R.A.G. appreciates the support of the Universidad Nacional Autónoma de México, through a DGAPA Doctoral Fellowship.

# Table 1

## Observation Log

| Object | Filter | $\lambda_{eff}$ | FWHM | Exposure | Telescope |
|---|---|---|---|---|---|
| M99 | $g$ | 5000Å | 830Å | 40 min | Lick 1-m |
|  | $r_S$ | 6800Å | 1330Å | 35 min |  |
|  | $i$ | 7800Å | 1420Å | 40 min |  |
|  | $J$ | 1.25$\mu$m | 0.29$\mu$m | 56 min | Kitt Peak 1-m |
|  | $K'$ | 2.16$\mu$m | 0.33$\mu$m | 56 min |  |
| M51 | $K'$ | 2.16$\mu$m | 0.33$\mu$m | 22 min |  |